\documentclass[runningheads]{llncs}
\usepackage[T1]{fontenc}
\usepackage{graphicx}
\usepackage{booktabs}
\usepackage{comment}
\usepackage{tabularray}
\usepackage{xcolor}
\usepackage{colortbl}
\usepackage{graphicx}
\usepackage{siunitx}
\usepackage{multirow}
\usepackage{amsmath}
\usepackage{amssymb}
\usepackage{array}
\usepackage[misc]{ifsym}
\newcommand{\corr}{(\Letter)}
\usepackage{mwe}

\begin{document}

\title{A Padding Method for Enhanced Encoding of Inorganic Structures with Varying Chemical Compositions}

\titlerunning{Enhanced Encoding of Inorganic Structures}

\author{Author information scrubbed for double-blind reviewing}

\author{Thang Dang\inst{1} \corr \and
Haderbache Amir\inst{1}  \and
Tzanakakis Alexandros\inst{1,2} \and
Yoshimoto Yuta \inst{1}}

\authorrunning{Thang Dang et al.}

\institute{Fujitsu Limited, Japan \email{\{thang.dang,haderbache.amir,fj5321it,yoshimoto.yuta\}@fujitsu.com}
\and
National Technical University of Athens, Greece \email{Alexandros@ntua.gr}}

\maketitle  

\begin{abstract}
Designing novel inorganic materials through generative models remains an important challenge for material science, driven by the complexity and diversity of inorganic structures across expansive chemical compositions and structural landscape. The vast combinatorial space of inorganic compounds demands innovative, AI-driven approaches to overcome limitations in generative accuracy and efficiency. To address this, we introduce a novel method that redefines the encoding and generation of inorganic materials by utilizing domain-specific symmetry-aware representation. Our approach not only refines the representation of intricate inorganic structures but also contributes to the field of material discovery by enhancing the precision and stability of generated candidates.
Central to our methodology is a novel padding technique that exploits crystal symmetry information to enhance the encoding process. By integrating Wyckoff position length-aware padding into an encoder architecture, we achieve a more robust informed representation of inorganic materials. This symmetry-driven enhancement improves deep learning models to generate stable, previously unexplored inorganic structures with superior accuracy and computational efficiency. Furthermore, we introduce an end-to-end system that leverages the machine learning potential models to seamlessly generate novel, even those unseen in the training data, and stable inorganic materials from initial data to validated output. This pipeline integrates advanced generative models with stability analysis, marking a significant leap forward in the automated exploration and design of next-generation inorganic materials. Our method improved reconstruction accuracy 5.3\% in proton conductor data, and generated 63.5\% more novel stable inorganic material to baseline model on the perov-5 dataset.

\end{abstract}

\section{Introduction}
The discovery and design of novel inorganic materials~\cite{thakur2024review} is an important research direction for material science, underpinning advancements in fields ranging from energy storage~\cite{liu2024recent} and catalysis~\cite{karpovich2023interpretable,hoskuldsson2023high} to electronics~\cite{qiao2023soft} and proton conductor for fuel cell~\cite{szaro2024first}. Inorganic materials exhibit a vast and intricate structural diversity, characterized by complex chemical compositions and a wide array of crystal symmetries. This diversity, while offering immense potential for innovation, presents a difficult challenge: the combinatorial space of possible inorganic compounds is so expansive that traditional experimental and computational methods struggle to explore it efficiently. As a result, the development of new inorganic materials has often been constrained by the limitations of conventional approaches, which are time-consuming, computationally expensive, and unable to systematically navigate the full scope of possible structures.





In recent years, machine learning (ML) has emerged as an effective tool to accelerate materials discovery~\cite{zeni2023mattergen,10650978}, with generative models~\cite{xiecrystal,sakai2023self} showing particular promise in designing novel compounds. However, applying these models to inorganic materials remains a significant hurdle due to the challenges posed by their structural complexity and the need for precise, stable outputs that adhere to physical and chemical principles. Existing generative frameworks~\cite{zhu2024wycryst} often fall short in capturing the nuanced features of inorganic systems such as lattice symmetries and atomic coordination environments leading to inaccuracies in generated structures or computationally inefficient workflows. To enhance ML-based material design, there is an urgent need for innovative strategies that can both represent the rich diversity of inorganic structures and generate viable candidates with high fidelity.






In this work, we present a novel framework that redefines the encoding and generation of inorganic materials through a symmetry-aware approach. By introducing a novel padding technique that leverages crystal symmetry information to improve the conventional Wyckoff representation method~\cite{zhu2024wycryst}, specifically, by introducing a Wyckoff position length-aware padding, we achieve a more robust and informed representation of inorganic structures within a sophisticated encoder architecture. This advancement enables deep learning models to generate stable and previously unexplored inorganic compounds with unprecedented accuracy and efficiency. Beyond representation, our end-to-end pipeline integrates advanced generative models with stability analysis, creating a seamless process from initial material to validated design. This approach addresses limitations in generative accuracy while offering a pathway to automate the discovery of stable, next-generation inorganic materials, which could support advancements across diverse applications.



\textbf{Main Contributions.}

We propose a padding-based method, inspired by padding techniques in natural language processing (NLP)~\cite{gimenez2020semantic}, to preprocess inorganic data with varying chemical compositions relative to Wyckoff positions, enabling a novel approach for encoding complex inorganic structures. Unlike conventional method~\cite{zhu2024wycryst}, this technique accounts for the diversity in chemical elements and their spatial arrangements, facilitating the generation of robust representations. Furthermore, we investigate the generation of a wide variety of inorganic structures by leveraging latent space sampling that reflects the number of chemical compositions, allowing for the creation of diverse and previously unexplored inorganic candidates. To ensure practical applicability, we introduce an end-to-end system that evaluates the stability of these newly generated structures, seamlessly integrating design and validation. Consequently, our approach enables the production of a greater number of stable, unique, and diverse inorganic materials, outperforming existing methods under comparable configurations.

\begin{figure}[t]
\centering
\includegraphics[width=0.9\textwidth]{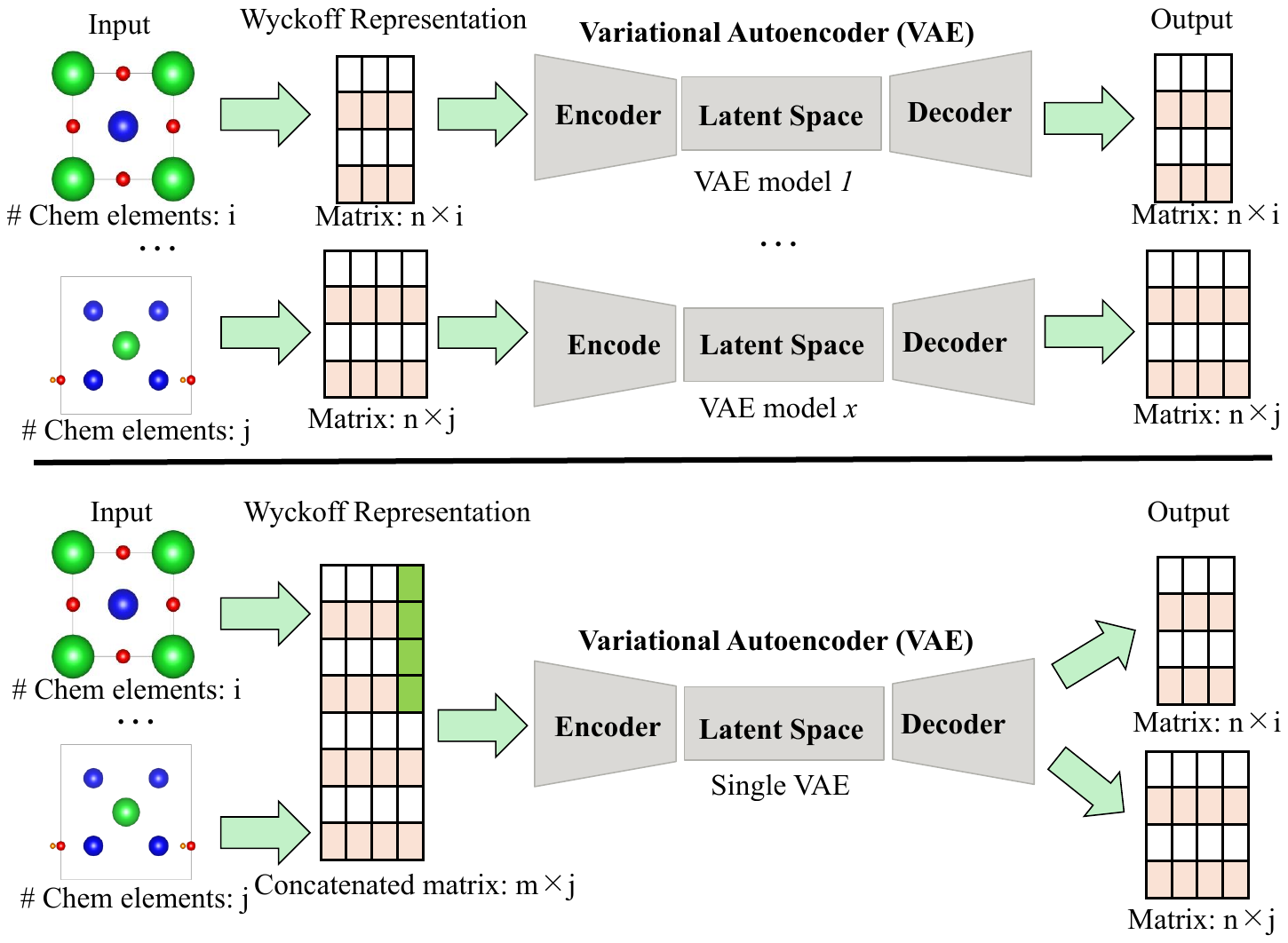}
\label{fig:VAE_flow}
\caption{Comparison of the conventional method (upper diagram) and the proposed method (lower diagram) for generating molecular data structures. Input: inorganic molecular structures; Output: Wyckoff representation matrix; \(n\): embedding dimension; \(i\): number of chemical elements.} 
\end{figure}


\section{Related Work}
Significant research efforts have focused on the generation of inorganic materials using advanced computational techniques. Generative model-based approaches, such as generative adversarial networks (GANs)~\cite{dan2020generative,fuhr2022deep}, variational autoencoders (VAEs)~\cite{han2023design,luo2024deep}, and diffusion models~\cite{yang2023scalable}, have garnered considerable attention for their ability to propose novel material compositions and structures. Another promising direction is inverse design~\cite{ren2022invertible,karpovich2024deep}, which enables the targeted generation of inorganic materials by specifying desired properties and working backward to identify suitable candidates. 
Large Language Model (LLM) based models~\cite{choudhary2024atomgpt,gruver2024finetuned} are also of research interest in generating inorganic generations along with taking advantage of the rapid progress in the field of generative AI.

Stability analysis is a critical step in validating newly generated inorganic materials. Several studies have employed ML-based methods, such as ML interatomic potentials derived from Graph Neural Networks (GNNs)~\cite{PhysRevLett.120.145301,reiser2022graph} or Gaussian process regression~\cite{deringer2021gaussian}, to assess thermodynamic stability and defect characteristics. Despite these advances, challenges persist. Capturing rare or complex defects such as grain boundaries or dislocations remains difficult due to limited training data. Additionally, interpreting ML outputs in a physically meaningful way continues to pose a significant hurdle, limiting the practical applicability of these models.

Among the studies most relevant to our research are the WyCryst method~\cite{zhu2024wycryst}, which introduce a Wyckoff position encoding scheme to train a VAE for reconstructing 3D crystal structures based on their symmetric properties. While innovative, this approach has a notable limitation: it struggles to accommodate Wyckoff position sequences of varying chemical composition lengths, which arise from differences in chemical elements, thus restricting its flexibility across diverse materials. Similarly, the SYMMCD method~\cite{levy2025symmcd} focuses on encoding asymmetric components within a unit cell and leverages crystal symmetry to reconstruct the full structure. However, this technique is constrained by its inability to generate novel inorganic materials with a broad range of chemical elements, limiting its applicability for exploratory material design. These shortcomings highlight the need for more adaptable and generative approaches, which our work seeks to address.

\section{Background}
Inorganic materials play a pivotal role in advancing technologies such as energy storage, catalysis, and electronics. The discovery of novel materials with tailored properties remains a cornerstone of materials science, yet the vast chemical and structural diversity of inorganic compounds presents significant challenges. Traditional experimental and computational approaches often struggle to efficiently explore this expansive design space. 
To contextualize our approach, this section reviews key concepts relevant to inorganic materials and the generation techniques.

\subsection{Inorganic material}
Inorganic materials are widely used in a wide range of substances like metals, ceramics, and oxides. These materials are prized for properties such as high melting points, electrical conductivity, or optical functionality, making them essential in applications like batteries, superconductors, and catalysts. Their characteristics stem from a crystalline framework defined by a lattice, a three dimensional, repeating array of points representing atomic positions. The smallest repeating piece of this lattice, called the unit cell, is shaped by three lengths \(a\), \(b\), \(c\) and angles \(\alpha\), \(\beta\), \(\gamma\).

The organization of atoms within and across unit cells is classified by a space group, a mathematical description combining symmetry operations like rotations, reflections, and translations with lattice types (e.g., cubic, hexagonal). Symmetry underpins this framework, governing how atoms are arranged identically under these operations, influencing properties like stability. Designing new inorganic materials requires understanding and manipulating these interconnected features lattice, unit cell, space group, and symmetry to achieve stable, functional structures with targeted properties.

\subsection{Wyckoff Positions}

Wyckoff positions are a fundamental concept in crystallography, describing the distinct symmetry sites within a crystal's unit cell as defined by its space group. Named after Ralph W. Wyckoff, these positions classify the possible locations of atoms based on the symmetry operations (e.g., rotations, reflections) of the crystal lattice. Each position is labeled with a letter (e.g., 2a, 2b) and corresponds to a specific multiplicity and set of coordinates, reflecting the degree of symmetry at that site. For example, a perovskites material such as BaZrO\(_3\) is defined by space group number 38, orthorhombic crystal system and wyckoff position as [``Ba'': [``2a''], ``Zr'': [``2b''], ``O'': [``2a'', ``4e'']]. In the context of inorganic material design, Wyckoff positions are critical for encoding structural information, as they capture the symmetry constraints that govern atomic arrangements. Incorporating this symmetry into deep learning models can enhance their ability to generate physically realistic crystal structures, a principle central to our proposed methodology.

\subsection{Energy Above Hull}

The energy above hull is a thermodynamic metric used to assess the stability of a material relative to its competing phases. In computational materials science, the convex hull represents the set of thermodynamically stable compounds at a given composition, plotted as formation energy versus composition. The energy above hull quantifies how far a material's formation energy lies above this baseline, with lower values indicating greater stability. For instance, an inorganic material has an energy above hull of 0 meV/atom lies on the hull and is stable, while one with a positive value is metastable or unstable, prone to decomposition into more favorable phases. This metric is widely used in material discovery to filter candidate structures generated by computational methods, including deep learning. A key challenge in generative modeling is ensuring that proposed inorganic materials exhibit low energy above hull values, reflecting realistic stability, a goal our approach aims to address through refined structural encoding.

\subsection{Variational Autoencoder}

Variational Autoencoders (VAEs) are a class of generative deep learning models that combine neural networks with probabilistic inference, widely applied in tasks like image synthesis and, more recently, material design. A VAE consists of an encoder, which compresses input data (e.g., crystal structures) into a latent space, and the decoder reconstructs the input data (at training phase) or can generate a new data from a sampled latent vector. Unlike traditional autoencoders, VAEs impose a probabilistic structure on the latent space, typically a Gaussian distribution, enabling smooth interpolation and sampling of new data points. In materials science, VAEs have been used to generate molecular structures or crystal lattices by learning patterns from databases like the Materials Project~\cite{jain2013commentary}. 
Our work builds on the Wyckoff VAE~\cite{zhu2024wycryst}, enhancing VAE performance through symmetry-aware techniques to produce novel, stable inorganic structures.


\section{Proposed Method}

Our proposed padding based method is shown in Fig.~\ref{fig:VAE_flow}, where the upper figure is the conventional method and the lower one is our method. A VAE framework is employed, comprising an encoder and a decoder, each designed with multiple interconnected layers to process and transform the input data effectively. The encoder compresses the input into a latent representation, while the decoder reconstructs the output from this compressed form. A distinguishing feature of the proposed methodology lies in the adaptation of the Wyckoff representation, which differentiates it from the related approach, also processed Wyckoff representation, but did not support varying chemical elements. Our method addresses variability in chemical elements across datasets by standardizing the Wyckoff matrix dimensions. This is achieved by appending "0" values to the Wyckoff matrix for material structures with fewer chemical elements, thereby ensuring uniformity in matrix size irrespective of the number of elements present in a given material. This standardization step is critical, as it enables consistent processing of diverse material datasets without introducing dimensional mismatches that could impair model performance.


By employing this approach, we eliminate the need to train multiple separate VAE models for different chemical compositions. Instead, a single VAE model can be trained on the entire dataset. This not only enhances computational efficiency but also enables the model to learn from the full diversity of the training data, capturing a broader range of structural and compositional patterns. As a result, the method not only simplifies the training process but also improves the model's ability to generate novel inorganic materials.


Our method processes the input data, which includes a chemical formula, space group number, and Wyckoff position dictionary, alongside an integer \( n_e \) defining the maximum number of chemical elements.
It comprises the following steps:
\begin{enumerate}
	\item \textbf{Initialization and Preprocessing.} Elemental embeddings are loaded as a matrix \( F_{\text{CGCNN}} \in \mathbb{R}^{|E| \times d} \), where \( d \) is the embedding dimension (we ultilize CGCNN model~\cite{PhysRevLett.120.145301} for embedding inorganic structures). For each element \( e_i \in E \), \( F_{\text{CGCNN}}[i-1, :] \) provides its feature vector.
	
	\item \textbf{Compositional Encoding.} For the \( x \)-th material in \( D \) (number of samples), the composition \( C_x \) is parsed to extract \( n_x \leq n_e \) elements. Atomic numbers \( Z_x = [z_1, z_2, \ldots, z_{n_x}] \) are mapped to a one-hot matrix \( O_x \in \{0, 1\}^{n_e \times |E|} \), where \( O_x[i, z_i-1] = 1 \) for \( i < n_x \), and padded with zeros for \( i \geq n_x \). Stoichiometric ratios \( R_x = [r_1, r_2, \ldots, r_{n_x}] \) are computed from \( C_x \), normalized by the total atom count \( n_{\text{atoms}}(C_x) \), and extended to \( \hat{R}_x \in \mathbb{R}^{n_e \times 1} \) with zeros for \( n_x < n_e \). CGCNN features form a matrix \( A_x \in \mathbb{R}^{n_e \times d} \), where \( A_x[i, :] = F_{\text{CGCNN}}[z_i-1, :] \) for \( i < n_x \), and zero otherwise.
	
	\item \textbf{Space Group Featurization.} The space group number \( s_x \in \{1, 2, \ldots, 230\} \) is encoded as a one-hot vector \( S_x \in \{0, 1\}^{230 \times 1} \), with \( S_x[s_x-1] = 1 \), representing the material’s crystallographic symmetry.
	
	\item \textbf{Wyckoff Position Featurization.} Wyckoff positions from \( D[x] \) are parsed into a matrix \( W_x \in \mathbb{R}^{n_e} \). For each element \( e_i \) and its Wyckoff label \( w_{ij} \) (e.g., ``4a''), the site index \( k \) (from 0\(-\)25: `a' to `z') increments \( W_x[i, k] \), and the multiplicity \( m_{ij} \) (e.g., 4) is stored in \( W_x[i, k+26] \). The final Wyckoff feature matrix is \( \hat{W}_x = W_x[:, :26] \in \mathbb{R}^{26 \times n_e} \), with cell ratios \( CR_x = \sum_{k=0}^{25} W_x[k, :] \cdot W_x[k+26, :] \in \mathbb{R}^{n_e} \). Outputs are lists \( W = [M_1, M_2, \ldots, M_N] \) and \( SG = [S_1, S_2, \ldots, S_N] \).
	
\end{enumerate}

The Wyckoff representation returns \( W \) (Wyckoff Position) and \( SG \) (Space Group), encoding atomic, compositional, and symmetry information for \( N \) materials. This representation, designed for inorganic systems, integrates Wyckoff positions and space groups to enhance VAE reconstruction and generation. By accommodating variable \( n_e \) as the longest wyckoff position in each batch and leveraging pre-trained embeddings, it overcomes prior limitations in compositional diversity.

\begin{figure}[h]
	\centering
	\includegraphics[width=0.8\textwidth]{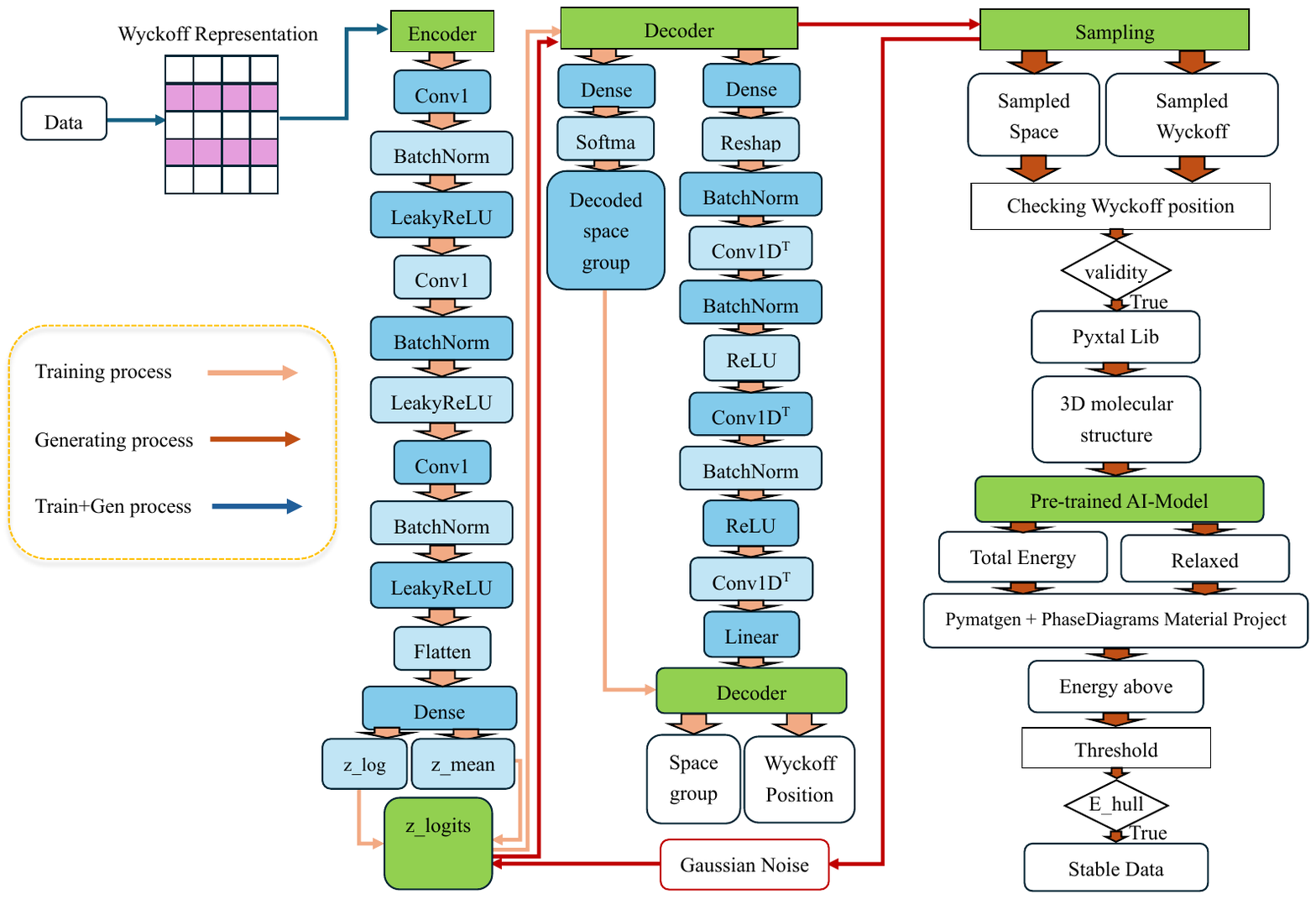}
	\caption{Our end-to-end system for generating new inorganic materials.}
	\label{fig:fullflow}
\end{figure}

Fig.~\ref{fig:fullflow} illustrates our end-to-end framework for integrating data preprocessing, generative modeling, structural validation, and stability analysis. The workflow proceeds through the following sequential steps:

\begin{enumerate}
	\item \textbf{Wyckoff Representation Conversion.} The process begins with input data comprising chemical compositions, space group numbers, and Wyckoff position dictionaries being transformed into a symmetry-aware Wyckoff representation. This step encodes atomic and crystallographic information into a fixed dimensional format suitable for machine learning, as detailed in previous section.
	
	\item \textbf{VAE Model Training.} The Wyckoff representations are used to train a single VAE model. The VAE learns a latent space distribution that captures the underlying patterns of the input data, enabling both reconstruction and generation of material structures. We employed four loss functions as: KL Divergence, Space Group Loss, Reconstruction Loss and Wyckoff Position Loss, are used during training phase.
	
	\item \textbf{Latent Space Sampling with Gaussian Noise.} To generate new candidate materials, we sample from the latent space of the trained VAE. A Gaussian noise function is applied to perturb the latent vectors, introducing controlled variability. The perturbed vectors are then decoded to produce new Wyckoff positions and corresponding space group encodings.
	
	\item \textbf{Wyckoff Position Validation.} The decoded Wyckoff positions are validated for physical consistency, ensuring they adhere to crystallographic rules (e.g., valid site occupancies and multiplicities). Invalid configurations are discarded at this stage.
	
	\item \textbf{3D Structure Generation with Pyxtal~\cite{pyxtal}.} Valid Wyckoff positions and space groups are converted into three-dimensional crystal structures using the Pyxtal library~\cite{pyxtal}. This step translates the abstract symmetry data into explicit atomic coordinates.
	
	\item \textbf{Structure Relaxation and Energy Prediction.} The generated 3D structures are relaxed using a pretrained machine learning potential (we used pretrained CHGNet~\cite{deng_2023_chgnet} or M3GNet~\cite{chen2022universal} models in this work). This relaxation refines atomic positions and lattice parameters, after which the model predicts the total energy of the relaxed structure.
	
	\item \textbf{Energy Above Hull Calculation.} Stability is assessed by calculating the energy above the convex hull~\cite{anelli2018generalized} (\( E_{Hull} \)) using pymatgen~\cite{ong2013python} and phase diagram data~\cite{Riebesell2023} from the Materials Project~\cite{jain2013commentary}. This metric quantifies the thermodynamic stability of the new structure relative to known stable phases.
	
	\item \textbf{Stability Screening with \( E_{Hull} \) Threshold.} Finally, a predefined \( E_{Hull} \) threshold (e.g., 0.08~eV/atom~\cite{miller2024flowmm},  0.1~eV/atom~\cite{gruver2024finetuned}, and 0.5~eV/atom) is applied to filter the generated materials. Structures with \( E_{Hull} \) below this threshold are classified as stable and retained as viable candidates for further analysis or experimental synthesis.
\end{enumerate}

This workflow seamlessly integrates generative modeling with structural and energetic validation, leveraging domain-specific representations and efficient machine learning potentials. By systematically screening for stability, our framework identifies promising new inorganic materials while minimizing computational overhead, as demonstrated in our experimental results (Section~\ref{sec:experiments}).

\section{Experiments}
\label{sec:experiments}
\subsection{Experimental Setup}
We evaluated our approach across multiple tasks using benchmark datasets tailored to the discovery of novel inorganic materials. Unlike prior studies~\cite{zhu2024wycryst,gruver2024finetuned}, which predominantly assessed the stability of newly generated materials while overlooking the diversity of their chemical compositions, our system explicitly addresses this gap. This emphasis on compositional diversity, alongside stability, represents a key innovation in our implementation.

\textbf{Benchmark datasets.} We used 3 datasets representing different number of chemical compositions. 1. \textbf{Perov-5}~\cite{castelli2012new} includes 18,928 perovskite materials that have the same structure but are different in chemical compositions. 2. \textbf{mp-20}~\cite{jain2013commentary} includes 45,231 inorganic samples from Materials Project. 3. \textbf{Proton-conductor}~\cite{szaro2024first} dataset has more than 4,793 data specifically designed as proton-conducting ceramic (PCC) electrolytes for solid-oxide fuel cells.

The implementation is structured into three core components:

\textbf{Preprocessing and Model Training.} A key focus of our research is the development of a novel padding technique for processing Wyckoff position sequences. For each encoded Wyckoff sequence, we append padding to shorter sequences, standardizing them into uniform batch blocks. This enhanced Wyckoff representation is then used to train a VAE model. In order to eliminate the influence of imbalance between data partitioned by chemical compositions, we only use data with balanced and sufficiently large amounts in each benchmark data set for training the VAE model. This is consistent with the research task of focusing on generating inorganic structures with diverse chemical compositions. Table~\ref{tab:data_distribution} shows the distribution of datasets (number of samples) used in our implementation, where x\_chem with x=[2, 3, 4, 5] indicates the number of chemical elements. In each training phase, we split the dataset into 80\% training and 20\% validation, with random shuffling applied. The hyperparameter settings are as follows: learning rate of \(2 \times 10^{-4}\), batch size of 256, 1,000 epochs, and the Root Mean Square Propagation (RMSprop) optimizer.

\begin{table}[h]
    \centering
    \caption{Distribution of data types across chemical composition counts (2\_chem to 5\_chem) for each dataset.}
    \label{tab:data_distribution}
    \begin{tabular}{l@{\extracolsep{10pt}}cccc}
        \toprule
        \textbf{Data Type} & \textbf{2\_chem} & \textbf{3\_chem} & \textbf{4\_chem} & \textbf{5\_chem} \\
        \midrule
        perov-5~\cite{castelli2012new} & NOT-USED & 5,512 & 10,660 & 2,652 \\
        mp-20~\cite{jain2013commentary} & 9,175 & 26,400 & NOT-USED & NOT-USED \\
        proton-conductor~\cite{szaro2024first} & NOT-USED & NOT-USED & 2,235 & 2,361 \\
        \bottomrule
    \end{tabular}
\end{table}

\textbf{Reconstruction and Accuracy Evaluation.} Reconstruction is the process that we use the decoder reconstructs the input Wyckoff representation at the training phase. We assess the trained VAE’s performance by measuring reconstruction accuracy, focusing on two critical parameters: Wyckoff positions and space groups. These elements are essential for accurately reconstructing inorganic material structures.

\textbf{Material Generation and Stability Analysis.} In the generation phase, we introduce Gaussian noise into the latent space before decoding to produce new inorganic materials that remain plausible and proximate to the training data. The generated structures are subsequently relaxed and evaluated for stability using two pre-trained models, CHGNet~\cite{deng_2023_chgnet} and M3GNet~\cite{chen2022universal}. To reduce computational expense, we adopt stability metrics from prior work~\cite{levy2025symmcd}, avoiding the need for Density Functional Theory (DFT)~\cite{schleder2019dft} calculations. For each implementation, after generating a large amount of inorganic data, we randomly selected 1,000 samples to check stability (same approach in~\cite{levy2025symmcd}).

\subsection{Experimental Results}
We assessed the reconstruction accuracy of our method against the Wyckoff VAE~\cite{zhu2024wycryst} baseline in predicting Wyckoff positions (Wyckoff Accuracy) and space group symmetries (SG Accuracy) across three datasets: perov-5, mp-20, and proton-conductor. The results, detailed in Table~\ref{tab:train_results}, demonstrate that our approach not only achieves competitive performance but also exhibits exceptional potential for generating accurate inorganic material structures across diverse chemical complexities.
\begin{table}[htbp]
	\centering
	\caption{Reconstruction Accuracy of Wyckoff position and Space Group (SG) predictions on validation datasets.}
	\label{tab:train_results}
	\setlength{\tabcolsep}{10pt}
	\begin{tabular}{lcc}
		\toprule
		\textbf{Method} & \textbf{Wyckoff Accuracy} & \textbf{SG Accuracy} \\
		\midrule
		\multicolumn{3}{l}{\textbf{perov-5 dataset}} \\
		\midrule
		Wyckoff VAE (3\_chem) & 99.8\% & 100\% \\
		Wyckoff VAE (4\_chem) & 99.9\% & 100\% \\
		Wyckoff VAE (5\_chem) & \textbf{100\%} & 100\% \\
		\textbf{Ours (3+4+5\_chem)} & 99.9\% & \textbf{100\%} \\
		\midrule
		\midrule
		\multicolumn{3}{l}{\textbf{mp-20 dataset}} \\
		\midrule
		Wyckoff VAE (2\_chem) & 92.8\% & 91.0\% \\
		Wyckoff VAE (3\_chem) & 93.4\% & 92.0\% \\
		\textbf{Ours (2+3\_chem)} & \textbf{94.8\%} & \textbf{92.80\%} \\
		\midrule
		\midrule
		\multicolumn{3}{l}{\textbf{proton-conductor dataset}} \\
		\midrule
		Wyckoff VAE (4\_chem) & 81.2\% & 92.80\% \\
		Wyckoff VAE (5\_chem) & 82.7\% & \textbf{96.80\%} \\
		\textbf{Ours (4+5\_chem)} & \textbf{88.0\%} & 91.10\% \\
		\bottomrule
	\end{tabular}
\end{table}
For the \textbf{perov-5 dataset}, Wyckoff VAE achieves near-perfect Wyckoff accuracies of 99.8\%, 99.9\%, and 100\% for 3, 4, and 5 chemical elements (3\_chem, 4\_chem, 5\_chem), respectively, with 100\% SG accuracy across all cases. Our method, applied uniformly to 3, 4, and 5 elements (3\_4\_5\_chem), delivers a Wyckoff accuracy of 99.9\% and SG accuracy of 100\%, matching or closely rivaling the baseline while simplifying the model to handle multiple complexities simultaneously an indicator of its robustness and versatility.

In the \textbf{mp-20 dataset}, Wyckoff VAE records Wyckoff accuracies of 92.8\% and 93.4\% for 2\_chem and 3\_chem systems, with SG accuracies of 91\% and 92\%, respectively. Our method (2\_3\_chem) significantly outperforms this, achieving 94.8\% Wyckoff accuracy and 92.8\% SG accuracy as an improvement of 1.4--2\% in Wyckoff prediction and up to 1.8\% in space group accuracy.

The \textbf{proton-conductor dataset}, with greater structural complexity, further underscores our method’s potential. Wyckoff VAE yields Wyckoff accuracies of 81.2\% and 82.7\% for 4\_chem and 5\_chem systems, with SG accuracies of 92.8\% and 96.8\%. In contrast, our method (4\_5\_chem) achieves a Wyckoff accuracy of 88\% and an SG accuracy of 91.1\%, surpassing the baseline by up to 5.3\% in Wyckoff prediction. While SG accuracy slightly dips below the 5\_chem baseline, our consistent improvement in Wyckoff accuracy across complex systems signals its promise for challenging applications.

We also evaluated the performance of our proposed method against the Wyckoff VAE baseline for generating stable inorganic materials, using the \(E_{Hull}\) (eV/atom) as a metric of stability. The results, summarized in Table~\ref{tab:gen_results}, report the number of generated materials with \(E_{Hull}\) values below thresholds of 0.08, 0.1, and 0.5 eV/atom, assessed using two state-of-the-art interatomic potential models: CHGNet and M3GNet. Our model consistently demonstrates superior performance compared to the Wyckoff VAE baseline across these datasets, particularly in identifying stable structures.

For the \textbf{perov-5 dataset}, stability checking by CHGNet, our method outperforms the Wyckoff VAE at lower thresholds (0.08 and 0.1 eV/atom), detecting more stable structures (e.g., 170 vs. 104 at 0.08 eV/atom with 3\_chem). At the 0.5 eV/atom threshold, our model remains competitive, often matching or slightly trailing Wyckoff VAE (e.g., 564 vs. 574 with 3\_chem). When evaluated with M3GNet, our model significantly excels, especially at 0.5 eV/atom (e.g., 371 vs. 275 with 3\_chem), highlighting its robustness in stability assessment.

In the \textbf{mp-20 dataset}, our model again shows strong performance. With CHGNet, it identifies more stable structures at all thresholds (e.g., 139 vs. 130 at 0.08 eV/atom with 2\_chem, and 550 vs. 362 with 3\_chem at 0.5 eV/atom). Paired with M3GNet, our model consistently outperforms Wyckoff VAE, notably at 0.5 eV/atom (e.g., 411 vs. 328 with 2\_chem), reinforcing its effectiveness in stability checking.

For the \textbf{proton-conductor dataset}, our model shines brightly. With CHGNet, it detects more stable structures across all thresholds (e.g., 272 vs. 232 at 0.08 eV/atom with 5\_chem). When paired with M3GNet, the improvement is dramatic, especially at 0.5 eV/atom (e.g., 366 vs. 26 with 4\_chem), showcasing its exceptional capability to assess stability under varying conditions.

\begin{figure}[h]
	\centering
	\begin{minipage}{0.48\textwidth}
		\centering
		\includegraphics[width=\textwidth]{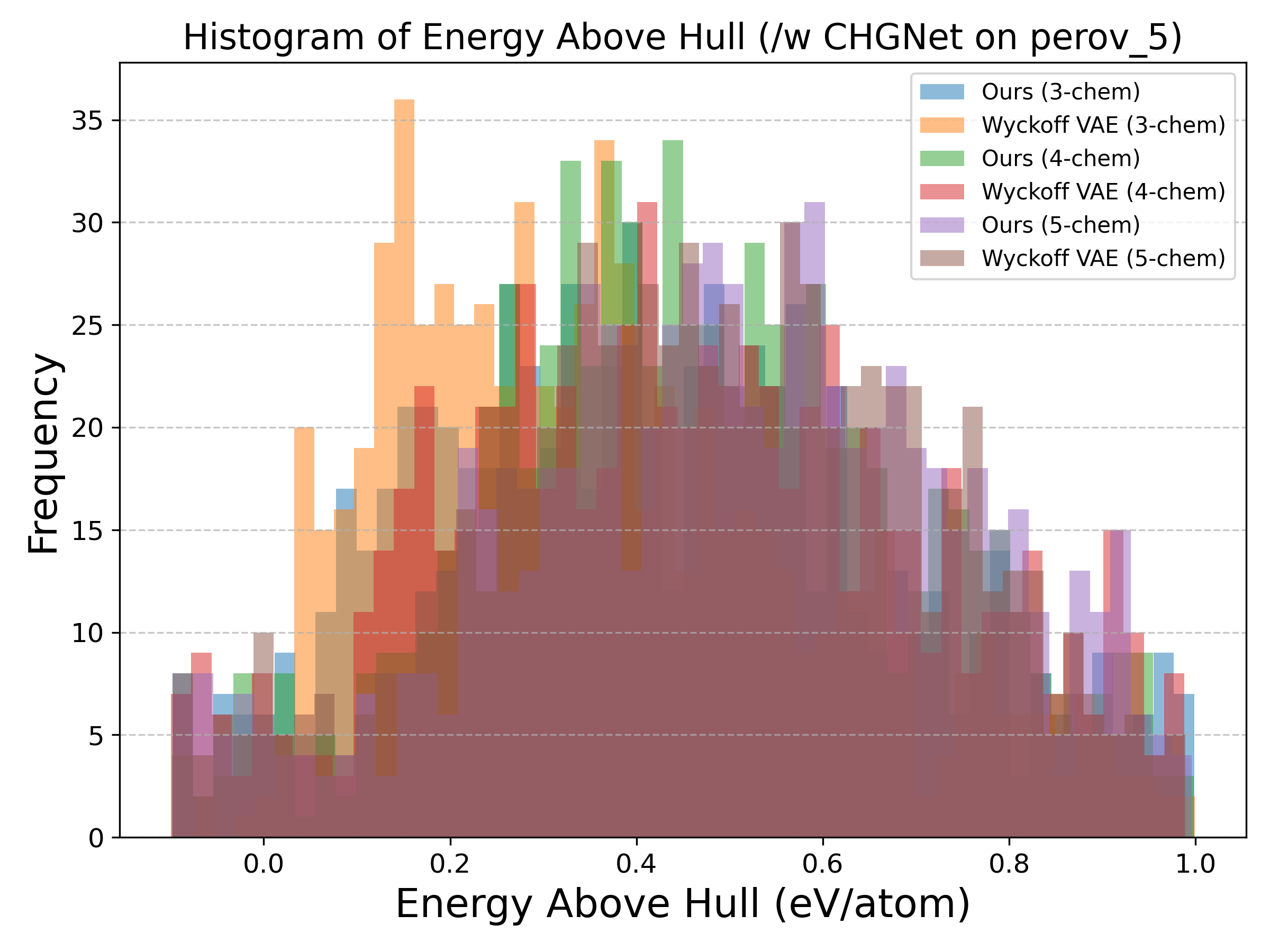}
		\caption{Histogram of new generated perov-5 data based on E\_hull by CHGNet.}
		\label{fig:hist_perov_chgnet}
	\end{minipage}
	\hfill
	\begin{minipage}{0.48\textwidth}
		\centering
		\includegraphics[width=\textwidth]{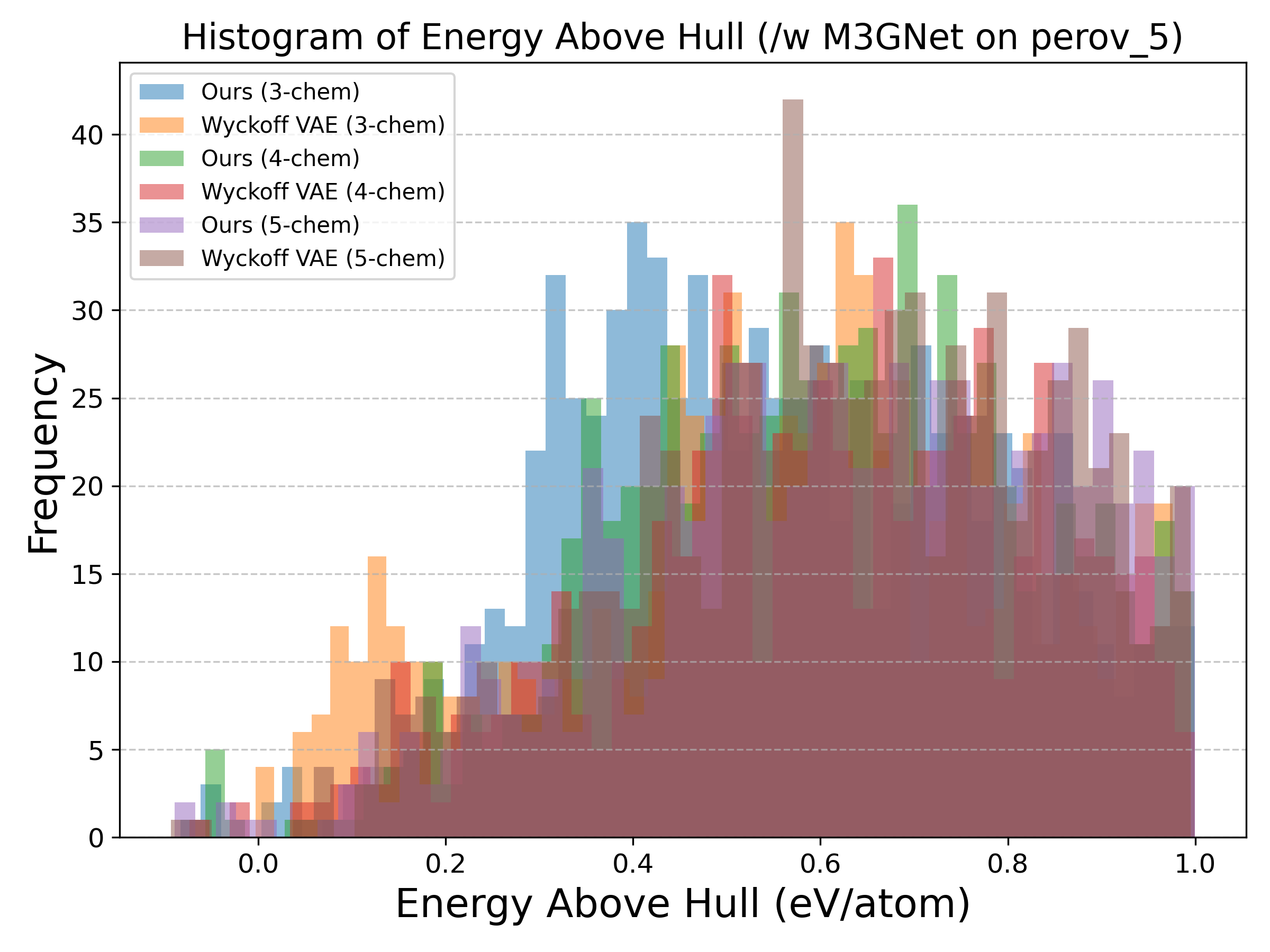}
		\caption{Histogram of new generated perov-5 data based on E\_hull by M3GNet.}
		\label{fig:hist_perov_m3gnet}
	\end{minipage}
\end{figure}

\begin{figure}[h]
	\centering
	\begin{minipage}{0.48\textwidth}
		\centering
		\includegraphics[width=\textwidth]{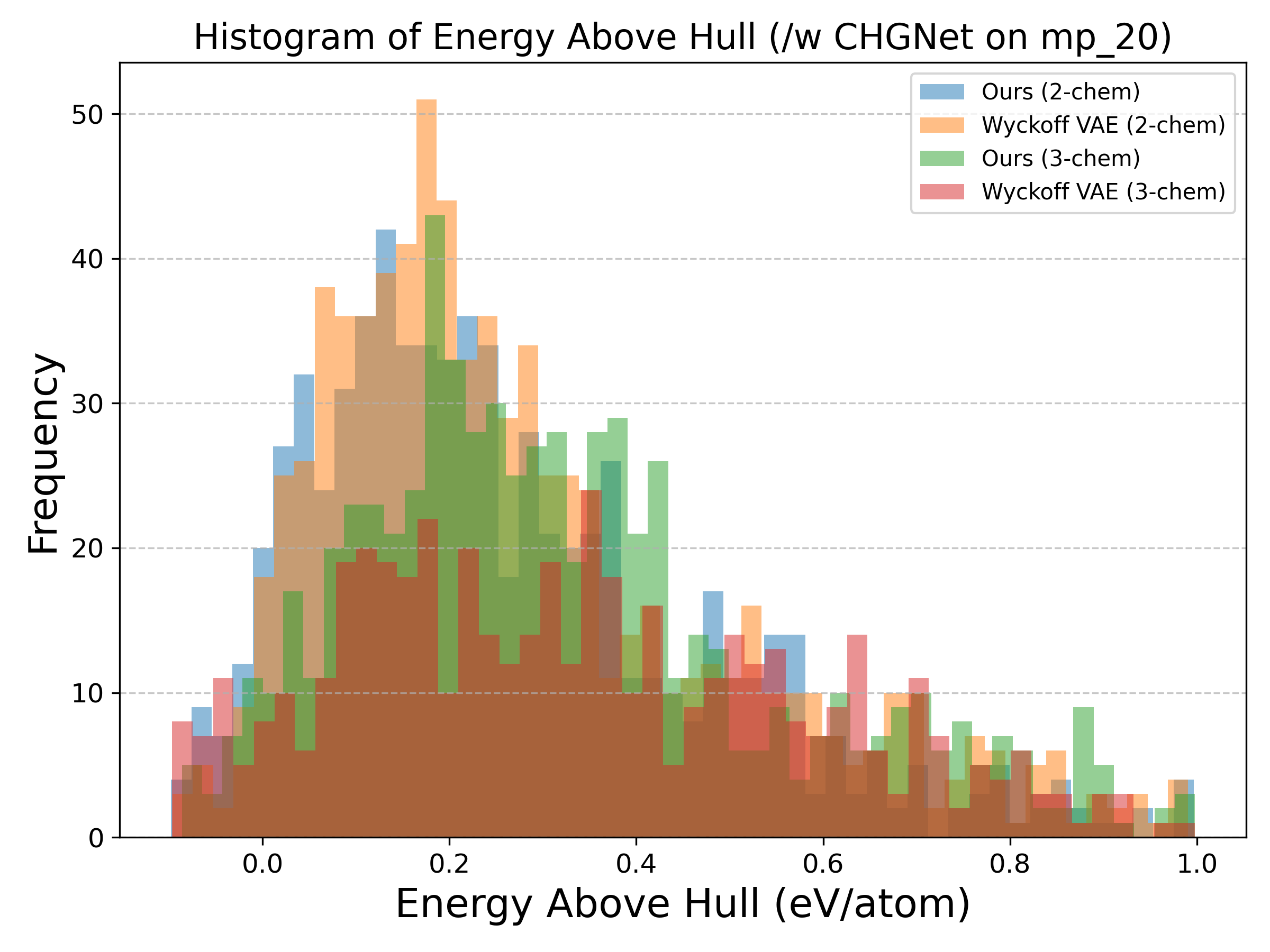}
		\caption{Histogram of new generated mp-20 data based on E\_hull by CHGNet.}
		\label{fig:hist_mp20_chgnet}
	\end{minipage}
	\hfill
	\begin{minipage}{0.48\textwidth}
		\centering
		\includegraphics[width=\textwidth]{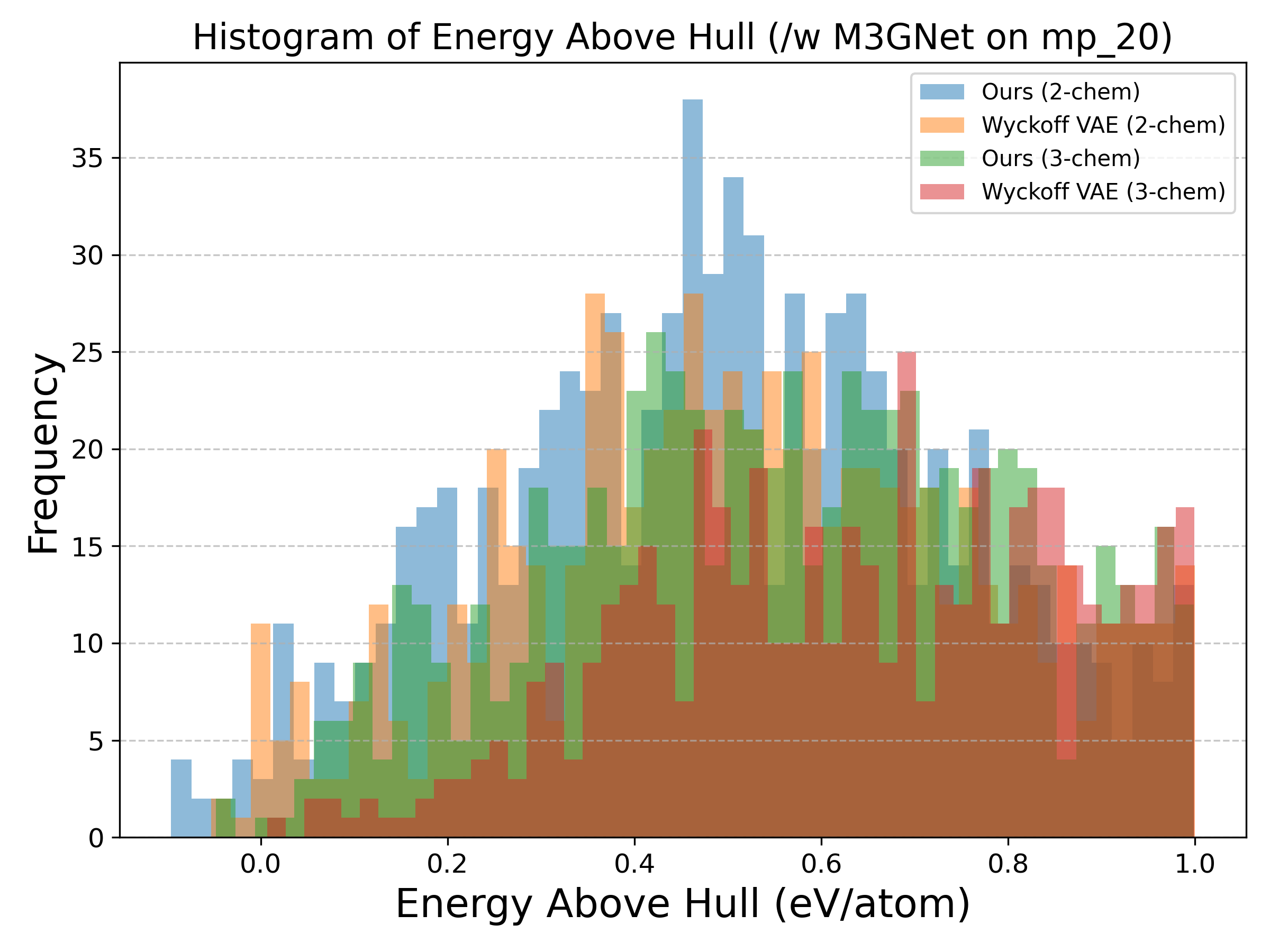}
		\caption{Histogram of new generated mp-20 data based on E\_hull by M3GNet.}
		\label{fig:hist_mp20_m3gnet}
	\end{minipage}
\end{figure}

\begin{figure}[h]
	\centering
	\begin{minipage}{0.48\textwidth}
		\centering
		\includegraphics[width=\textwidth]{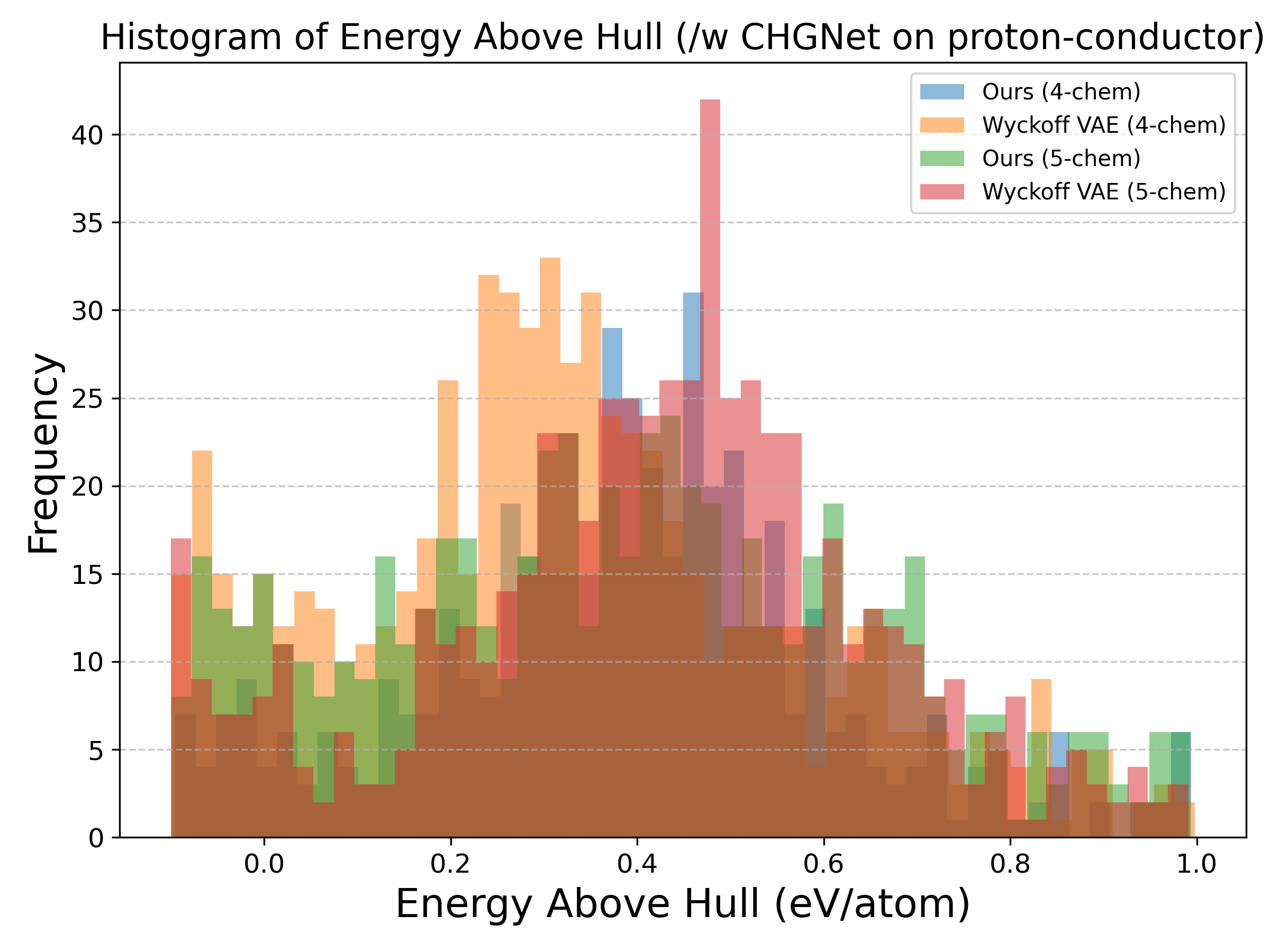}
		\caption{Histogram of new generated proton-conductor data based on E\_hull by CHGNet.}
		\label{fig:hist_proton_chgnet}
	\end{minipage}
	\hfill
	\begin{minipage}{0.48\textwidth}
		\centering
		\includegraphics[width=\textwidth]{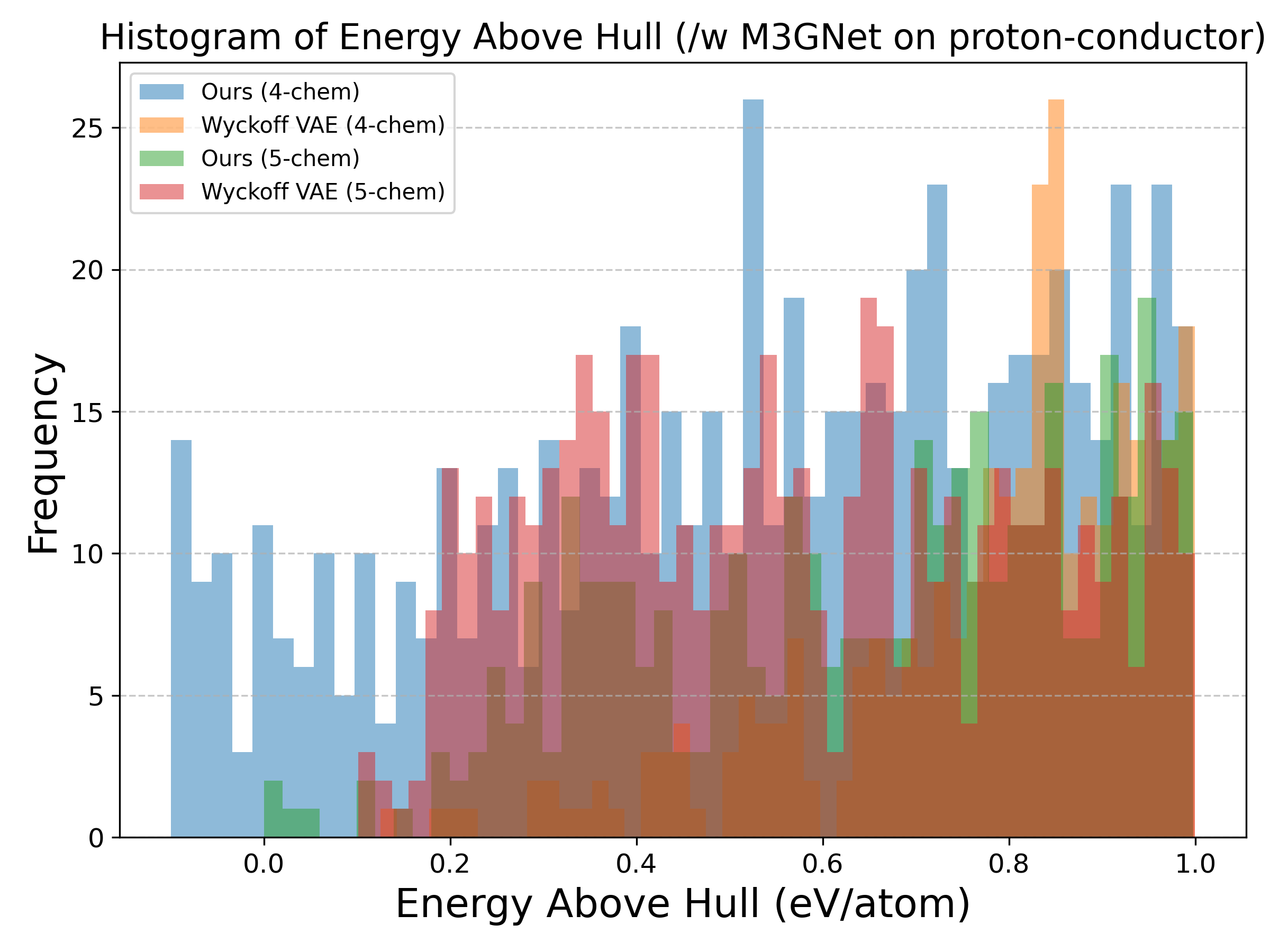}
		\caption{Histogram of new generated proton-conductor data based on E\_hull by M3GNet.}
		\label{fig:hist_proton_m3gnet}
	\end{minipage}
\end{figure}

Fig.~\ref{fig:hist_perov_chgnet}-\ref{fig:hist_proton_m3gnet} present histograms of newly generated materials based on $E_{\text{hull}}$ across three datasets, comparing the generalizability of our method with the Wyckoff VAE method.

Overall, our model excels in stability evaluation via $E_{\text{hull}}$ across all three datasets and thresholds (0.08, 0.1, and 0.5 eV/atom), consistently outperforming the Wyckoff VAE baseline, particularly when evaluated with CHGNet~\cite{levy2025symmcd}, making it a highly reliable tool for stability analysis.

\begin{table}[htbp]
	\centering
	\caption{Number of stable inorganic structures generated under three \(E_{Hull}\) thresholds (eV/atom) evaluated on perov-5, mp-20 and proton-conductor datasets. Noting that in each dataset, our method trained a single model on the entire training data, but in the generation phase, our framework generated material with varying chemical elements.}
	\label{tab:gen_results}
        \setlength{\tabcolsep}{3pt}
	\begin{tabular}{lccc}
		\toprule
		
		\textbf{Method} & \textbf{0.08 eV/atom} & \textbf{0.1 eV/atom} & \textbf{0.5 eV/atom} \\
		\midrule
		\midrule
		\multicolumn{4}{l}{\textbf{perov-5 dataset}} \\
		
		\midrule
		Wyckoff VAE (3\_chem) + CHGNet & 104 & 122 & \textbf{574} \\
		\textbf{Ours (3\_chem) + CHGNet} & \textbf{170} & \textbf{184} & 564 \\
		Wyckoff VAE (4\_chem) + CHGNet & 144 & 151 & 513 \\
		\textbf{Ours (4\_chem) + CHGNet} & \textbf{158} & \textbf{160} & \textbf{544} \\
		Wyckoff VAE (5\_chem) + CHGNet & 191 & 197 & \textbf{522} \\
		\textbf{Ours (5\_chem) + CHGNet} & \textbf{198} & \textbf{200} & 500 \\
		\midrule
		Wyckoff VAE (3\_chem) + M3GNet & 19 & \textbf{29} & 275 \\
		\textbf{Ours (3\_chem) + M3GNet} & \textbf{23} & 24 & \textbf{371} \\
		Wyckoff VAE (4\_chem) + M3GNet & 11 & 13 & 204 \\
		\textbf{Ours (4\_chem) + M3GNet} & \textbf{24} & \textbf{25} & \textbf{266} \\
		Wyckoff VAE (5\_chem) + M3GNet & 9 & 9 & 218 \\
		\textbf{Ours (5\_chem) + M3GNet} & \textbf{12} & \textbf{15} & \textbf{203} \\
		\midrule
		\midrule
		\multicolumn{4}{l}{\textbf{mp-20 dataset}} \\
		
		\midrule
		Wyckoff VAE (2\_chem) + CHGNet & 130 & 163 & \textbf{659} \\
		\textbf{Ours (2\_chem) + CHGNet} & \textbf{139} & \textbf{167} & 609 \\
		Wyckoff VAE (3\_chem) + CHGNet & 67 & 83 & 362 \\
		\textbf{Ours (3\_chem) + CHGNet} & \textbf{77} & \textbf{98} & \textbf{550} \\
		\midrule
		Wyckoff VAE (2\_chem) + M3GNet & 31 & 36 & 328 \\
		\textbf{Ours (2\_chem) + M3GNet} & \textbf{39} & \textbf{46} & \textbf{411} \\
		Wyckoff VAE (3\_chem) + M3GNet & 5 & 5 & 154 \\
		\textbf{Ours (3\_chem) + M3GNet} & \textbf{13} & \textbf{19} & \textbf{289} \\
		\midrule
		\midrule
		\multicolumn{4}{l}{\textbf{proton-conductor dataset}} \\
		\midrule
		Wyckoff VAE (4\_chem) + CHGNet & 168 & 174 & \textbf{511} \\
		\textbf{Ours (4\_chem) + CHGNet} & \textbf{178} & \textbf{182} & 468 \\
		Wyckoff VAE (5\_chem) + CHGNet & 232 & 237 & 570 \\
		\textbf{Ours (5\_chem) + CHGNet} & \textbf{272} & \textbf{280} & \textbf{581} \\
		\midrule
		Wyckoff VAE (4\_chem) + M3GNet & 1 & 1 & 26 \\
		\textbf{Ours (4\_chem) + M3GNet} & \textbf{162} & \textbf{167} & \textbf{366} \\
		Wyckoff VAE (5\_chem) + M3GNet & 0 & 0 & \textbf{228} \\
		\textbf{Ours (5\_chem) + M3GNet} & \textbf{7} & \textbf{8} & 108 \\
		\bottomrule
	\end{tabular}
\end{table}

\section{Conclusion}

The proposed method introduces a novel padding technique to create a unified Wyckoff representation for training data encompassing diverse chemical compositions.
Our method standardizes matrix wyckoff dimensions across datasets, enabling integration of structures with varying chemical elements (e.g., 2 to 5 elements). This approach eliminates the need for multiple, composition-specific VAE models, allowing a single VAE to be trained on the entire dataset.

As demonstrated in our implementation results, this unified framework significantly enhances both reconstruction and generation. Across the perov-5, mp-20, and proton-conductor datasets, our method achieves Wyckoff reconstruction accuracies of up to 99.9\%, 94.8\%, and 88\%, respectively, and SG accuracies reaching 100\%, 92.8\%, and 91.1\%, consistently matching or surpassing the Wyckoff VAE baseline. By learning from the full diversity of the training data, our model captures a wide range of structural and compositional patterns, resulting in improved reconstruction accuracy most notably a 5.3\% gain in Wyckoff accuracy for complex proton-conductor systems and a 1.4--2\% improvement in the mp-20 dataset. 

In the generation task, our method demonstrates exceptional performance by applying varying \(E_{Hull}\) thresholds of 0.08, 0.1, and 0.5 eV/atom and leveraging pretrained CHGNet and M3GNet models. Across three benchmark datasets, our approach consistently outperforms the Wyckoff VAE baseline, generating a greater number of stable inorganic materials. For instance on perov-5 data, with CHGNet, our method achieves up to 63.5\% more stable structures at 0.08 eV/atom for 3\_chem, while with M3GNet, it yields a 34.9\% increase at 0.5 eV/atom for 3\_chem.
These outcomes affirm the remarkable potential of our method as a versatile and powerful tool for accelerating the discovery of novel inorganic materials with practical synthesis viability.

\begin{credits}
\subsubsection{\ackname}
Hidden for double-blind review.
\subsubsection{\discintname}

\end{credits}
%
%
%
\bibliographystyle{splncs04}
\bibliography{mybibliography}
%

\end{document}